\documentclass{article}

\usepackage{PRIMEarxiv}

\usepackage[utf8]{inputenc} 
\usepackage[T1]{fontenc}    
\usepackage{hyperref}       
\usepackage{url}            
\usepackage{booktabs}       
\usepackage{amsfonts}       
\usepackage{nicefrac}       
\usepackage{microtype}      
\usepackage{lipsum}
\usepackage{fancyhdr}       
\usepackage{graphicx}       
\usepackage[utf8]{inputenc} 
\usepackage{amssymb,amsmath,amsthm} 
\graphicspath{{media/}}     

\title{Thermal diode assisted by geometry under cycling temperature}

\author{
  L. Zurdo \\
  Laboratorio de Modelado y Simulación Computacional \\
  Instituto de Ciencias, Universidad Nacional de General Sarmiento\\
  Los Polvorines, Buenos Aires.\\
  Corresponding author.\\
  luislucianozurdo@gmail.com \\
   \And
  L Chej, A. Monastra, and F. Carusela \\
  Laboratorio de Modelado y Simulación Computacional \\
  Instituto de Ciencias, Universidad Nacional de General Sarmiento\\
  Los Polvorines, Buenos Aires.\\
  Consejo Nacional de Investigaciones Científicas y Técnicas, Argentina.\\
}

\begin{document}
\maketitle

\begin{abstract}
Technological progress in electronics usually requires their use in increasingly aggressive environments, such as rapid thermal cycling and high power density. Thermal diodes appear as excellent candidates to thermally protect critical electronic components and ensure durability and reliability. We model the heat transport across a square plate with a hole subjected to an oscillating external temperature, such spatial and temporal symmetries are broken. We find rectification of the heat current that strongly depends on the frequency and the geometry of the hole. This system behaves as a thermal diode that could be used as part of a thermal architecture to dissipate heat under cycling temperature conditions.
\end{abstract}

\keywords{Heat transport \and Thermal diode \and Thermal cycling \and Thermal rectification}

\section{Introduction}
Many types of systems or devices operate under thermal conditions that entail a rapid redirection of the heat flux or its intensity \cite{heThermalManagementTemperature2021}.  
Examples are sensors \cite{draghici400SensorBased2019}, different kinds of electronic and energy conversion devices \cite{nguyenActiveThermalCloak2015} as well as some emerging technologies that operate under extreme natural or artificial environments. This is the case of temperature cycling operative conditions as in  the case of space, nuclear, oil industries, or harsh weather conditions. Just to give some few interesting examples we can mention electronic devices, which at present are essentially built in ultra-compact and extreme packaging designs \cite{hombergerSiliconcarbideSiCSemiconductor2004}. This increases substantially the thermal loads and consequently requires a specific thermal management for reliable operation. Examples of harsh cycling conditions are found in the spatial and construction industries. Low orbit satellites are exposed to a 90 minutes temperature period going from $-100^o$C up to $100^o$C when it goes from dark shadow to direct sun irradiation \cite{thirskSpaceflightEnvironmentInternational2009}. Regarding the construction industry in regions where there are extreme daily changing temperatures, it is required to preserve a stable temperature in the interiors with an energy consumption as low as possible \cite{al-tamimiEffectGeometryHoles2017}.

Thermal rectifiers can be suitable candidates for an efficient and intelligent thermal management. A thermal rectifier or thermal diode is a device based on a physical phenomenon to manipulate and control the direction of the heat flow, and eventually its intensity \cite{venkateshwarInfluenceNaturalConvection2021}.  There are passive or active thermal diodes to control or rectify the heat current that underlie  on different physical mechanisms. Passive thermal diodes are usually based on fluids or solid materials whose properties can be managed without the need of power sources \cite{kasprzakHightemperatureSiliconThermal2020}. On the other hand, active thermal diodes requires the application of an external energy source \cite{zhuTemperatureGatedThermalRectifier2014}.

There are several strategies to achieve large rectification ratios. Some studies \cite{wongReviewStateArt2021,cottrillPersistentEnergyHarvesting2019, wangExperimentalStudyThermal2017, kasprzakHightemperatureSiliconThermal2020} show that materials with different geometric modifications and patterings, such as nano-holes, deposition of nanoparticles or thin films, 
asymmetric graphene ribbons, or mass graded systems  are excellent candidates to build the necessary asymmetry to generate the thermal rectification phenomenon \cite{yangThermalRectificationNegative2007}, \cite{yangThermalRectificationAsymmetric2009}. It is also possible to increase the rectification ratio by a series circuit of thermal rectifiers \cite{huSeriesCircuitThermal2017}
These materials and devices are ideal platforms as components of systems that operate under thermal cycling.  

In this work we present a dynamical thermal rectifier made of a solid plate with a hole. Our primary interest is to understand how we can manage the heat flow by tuning the hole geometry, size and the  frequency of the temperature cycling, in order to improve the thermal diode effect. In particular, we study numerically the local and global rectification performance finding a resonant phenomenon caused by a synergy between the geometry and the thermal cycling. Finally we discuss potential applications.

\section{The Model}

\begin{figure}[ht]
    \centering
    \includegraphics[scale=0.6]{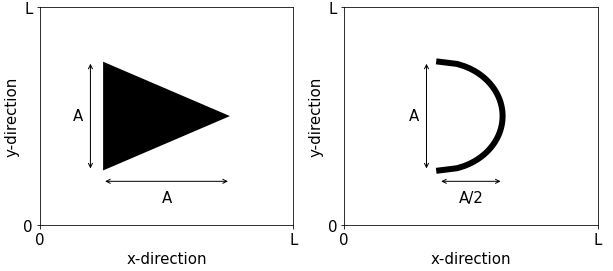}
    \caption{Schematic representation of a square plate of side $L$ with two different hole geometries. The triangle has a base and a height equal to $A$. The crescent is a semicircle of diameter $A$ and thickness $0.025L$ (which is kept constant for different diameters). Both holes are located at equal distance from the left and right walls of the domain. }
    \label{fig:holesSize}
\end{figure}

We model a solid square plate of side $L$, with a hole in its interior. We study two different hole shapes: i) a triangle with the same base and height $A$; ii) a semicircular crescent of diameter $A$ and width 0.025 $L$, see Fig. \ref{fig:holesSize}. The holes are also varied in size, to study how they affect the heat flow through the plate. We define a cross-section parameter $C = A / L$, taking three different values: 0.25, 0.50 and 0.75.

We consider a material of high thermal conductivity, such that loses due to radiation and convection over and below the plate can be neglected. This implies an effective 2D model. 

The evolution of the temperature field $T$ as a function of position $(x,y)$ and time $t$ inside the plate is governed by the heat equation
\begin{equation}
   \frac{\partial T}{\partial t} =  \alpha \left( \frac{\partial^2 T}{\partial x^2} +  \frac{\partial^2 T}{\partial y^2} \right)
    \label{heatEquation}
\end{equation}

where $\alpha = \kappa / ( \rho c_p )$ is the thermal diffusivity of the material in terms of its thermal conductivity $\kappa$, specific heat $c_p$, and density $\rho$.

\begin{figure}[ht!]
    \centering
    \includegraphics[width =0.5\textwidth]{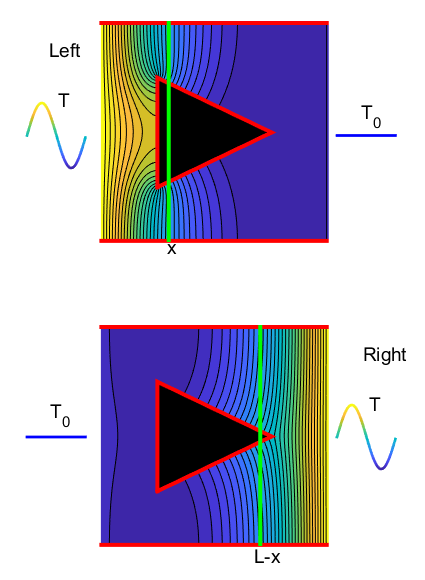}
    \caption{Snapshot of the temperature map for a {\it left-case} (top) and {\it right-case} (bottom) of position of temperature gradient where the yellow color indicates a hot spot and the blue color indicates a cold spot. The vertical green lines indicate the position where the heat current amplitude ${\cal J}$ is computed, Eq. (\ref{stationaryIntegralHeatFlux}). {\it Left} and {\it right} cases show on which side the temperature oscillates, while it remains fixed on the opposite side. These two current amplitudes are compared to compute the rectification coefficient. The red lines in both figures indicates the insulated boundaries and the black zone indicates a void zone.}
    \label{fig:rectification}
\end{figure}

In the borders of the hole and at the two parallel boundaries of the plate (see red lines in Fig. \ref{fig:rectification}) we impose an insulating condition (zero heat flux), that by the Fourier's law implies
\begin{equation}
    \frac{\partial T}{\partial \hat{n}} = 0
    \label{borderEquation}
\end{equation}
where $\hat {n}$ is the normal direction to such border. On the other lateral boundaries we consider a fixed $T_0$ on the left (right) side, and an oscillating temperature with a frequency $\omega$ and thermal amplitude $\Delta T$ on the right (left) side.
\begin{equation}    
T(x=L,y,t) = T_0
\label{constantTemperature}
\end{equation} 
\begin{equation}
T(x=0,y,t) = T_0 + \Delta T \cos \left({\omega t }\right)
\label{oscilatingTemperature}
\end{equation}
These boundary conditions model a device exposed to a cycling temperature around some average value, while on the other side there is a large heat bath at an approximately constant temperature. This configuration can be achieved in an experimental device using a modulated external electrical current applied on a metallic heater in contact with the plate, analogous to an electronic device operating under the action of an oscillating power. This can also be easily obtained by on-off controllers, with an on/off frequency congruent to the main one.

In previous works (e.g. \cite{zhangReversalThermalRectification2009, cottrillUltrahighThermalEffusivity2018, wehmeyerThermalDiodesRegulators2017}) it was shown that the action of external fields can break a time symmetry, resulting in a rectification of the heat flux. Another possibility is to produce a rectification from a spatial symmetry breaking \cite{lohThermalRectificationReversal2012, saeidNaturalConvectionPorous2005, liuBriefReviewRecent2019, zhaoReviewThermalRectification2022}. In a 2D solid uniform plate this can be made drilling holes with different geometries.

To simplify the model, we consider that the thermal properties of the material remain constant in the range $(T_0 - \Delta T, T_0 + \Delta T)$. The oscillating temperature in one border induces, even in a stationary regime, an oscillating heat flux in the bulk of the plate given by the Fourier's law

\begin{equation}
\vec{\mathbf{q}} (x,y,t)=-\kappa \Vec{\nabla T}
\label{fourierLaw}
\end{equation}

Considering $\alpha$, $\kappa$, $L$, and $\Delta T$ as fixed parameters of the system, we can work with dimensionless units: $L$ is the unit of length, $\tau = L^2 / \alpha$ is the unit of time, therefore $\Omega = 2 \pi / \tau$ the unit of angular frequency. The unit of heat flux is $q^* = \kappa \Delta T / L$ and the unit of heat current is  $J^* = q^* L =  \kappa \Delta T$. The value of $T_0$ is only a reference temperature, unimportant for the model, although the minimum temperature $(T_0 - \Delta T)$ should be high enough to avoid quantum effects or large variations in the thermal properties of the material. Regarding the size, the model is valid for plates of several centimeters up to micrometers, as far as $L$ is much bigger than the phonon mean free path.

We solve the equations numerically using the finite differences method \cite{ozisikFiniteDifferenceMethods2017, bergmanFundamentalsHeatMass2011}. 
We vary $\omega$ from $5.86 \Omega$ up to $586.45 \Omega$. The initial conditions for our simulation is a constant temperature of the plate. We have checked that after a time much larger than $\tau$ the stationary regime is achieved, where the temperature and the heat flux in every point of the plate oscillates with the same frequency than $\omega$, with a constant amplitude and a phase shift depending on the point.
\begin{eqnarray}
    T (x,y,t) &=& {\cal T}  (x,y) \cos (\omega t - \phi (x,y)) 
    \label{ecuacion_temperatura_estacionario_x_y}
    \\
    \vec{\mathbf{q}} (x,y,t) &=& \mathbf{{\cal Q}}  (x,y) \cos (\omega t - \varphi (x,y))
    \label{stationaryFluxEquation}
\end{eqnarray}
The amplitudes ${\cal T}$ and $\mathbf{{\cal Q}}$ of these fields, and their phase-shifts $\phi$ and $\varphi$, will strongly depend on the shape and size of the hole, as well as the considered position $(x,y)$.

To simplify the comparisons, we compute a heat current along a transversal section $x$ of the plate
\begin{equation}
    J(x,t) = \int_0^{L} q_x (x,y,t) \  \text{d} y
    \label{integralHeatlFlux}
\end{equation}
As well as for temperature and heat current fields, the heat current will also oscillate with the frequency $\omega$ in the stationary regime
\begin{equation}
    J(x,t) = \mathcal{J} (x) \cos( \omega t - \theta (x))
    \label{stationaryIntegralHeatFlux}
\end{equation}
Here ${\cal J} (x)$ is the current amplitude in position $x$, which can be computed by the maximum value it achieves in a period, while $\theta (x)$ is its phase-shift with respect to the forcing temperature. We compute its value at  ten different equidistant positions along the x-direction of the plate, as shown in the Fig.$\ref{fig:localHeatFluxes}$.

For each frequency, simulations were performed considering two different cases: an oscillatory temperature at the left boundary keeping constant the temperature on the opposite side, named as {\it left-case}, and the reverse situation named as {\it right-case}  as can be seen in the Fig. \ref{fig:rectification}.
This configuration exhibits asymmetric heat transfer characteristics along the $x$ axial direction.

The performance of the thermal diode is usually quantified by a thermal rectification coefficient \cite{xuSurfacePlasmonenhancedNearfield2018}. 
This factor is related to the difference between the current amplitude ${\cal J}$ at position $x$ in the {\it left} case, compared to the current amplitude in the {\it right} case computed at the mirror position $(L - x)$
\begin{equation}
    R(x) =\frac{ {\cal J}^R (x) - {\cal J}^L (L - x) }{ \text{max} (  {\cal J}^R (x) ,  {\cal J}^L (L - x) ) }
    \label{rectification}
\end{equation}
where the subscripts $L$ and $R$ correspond to the {\it left} and {\it right} cases schematized in the Fig. $\ref{fig:rectification}$.
Among several definitions for the rectification coefficient found in the literature, we have chosen the expression in Eq. \ref{rectification}, where the forward direction was arbitrarily chosen. This expression delimits $R(x)$ to the interval $[-1,1]$, so that $R(x)=0$ implies no rectification effect, while $R(x)=\pm 1$ implies a perfect thermal diode
\cite{wongReviewStateArt2021}.

\section{Results}

\subsubsection*{Local Heat Flux}

\begin{figure}[ht!]
    \centering
    \includegraphics[scale=0.6]{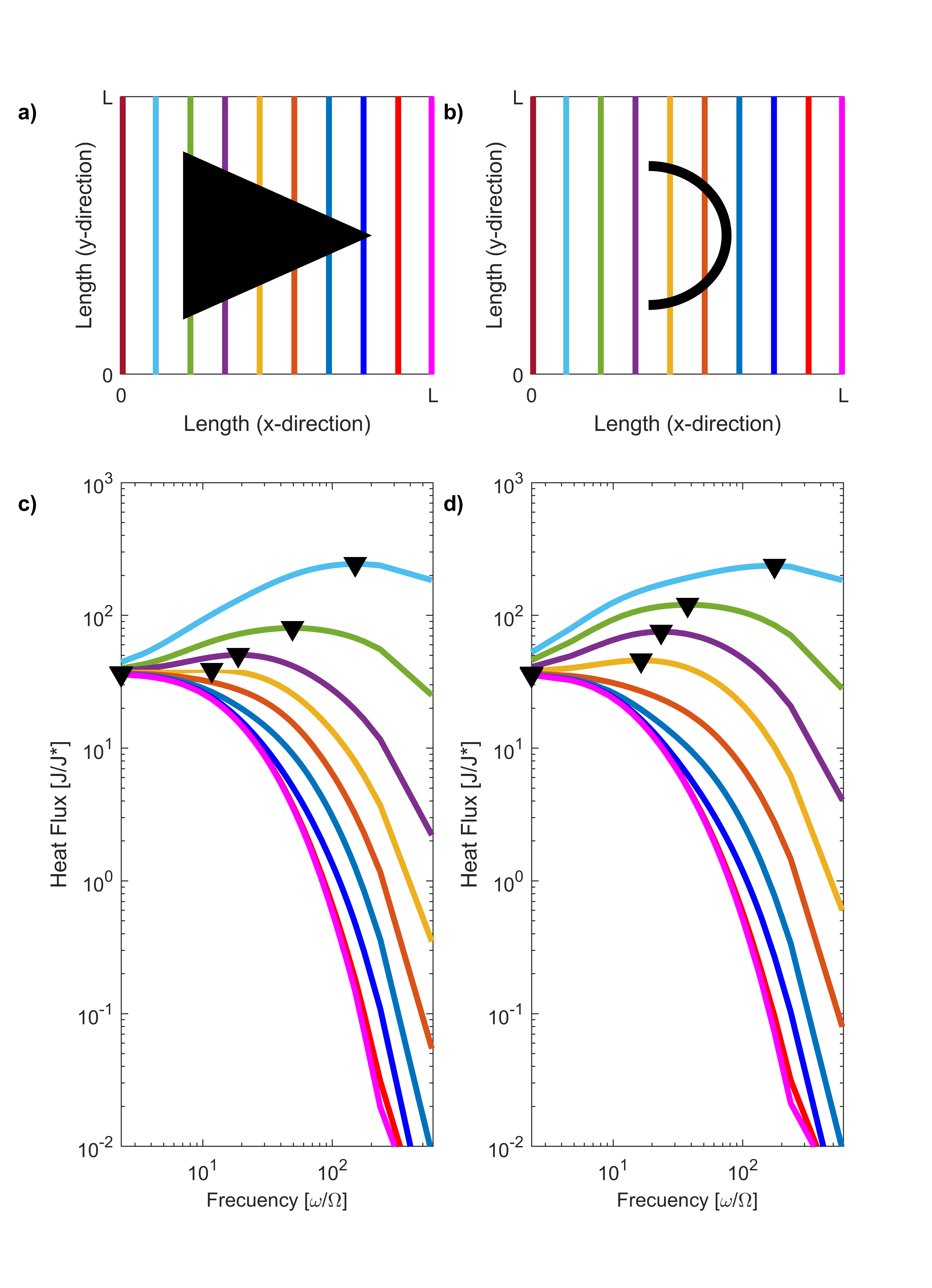}
    \caption{{
    Top panels: Color vertical lines show the cross-sections where the current amplitudes are computed for the triangular hole (a) and the crescent hole (b). Low panels: Heat current amplitude $\mathcal{J}(x)$ obtained by Eq. ($\ref{stationaryIntegralHeatFlux}$) as a function of the frequency $\omega$ for the triangular (c) and crescent(d) holes in the {\it left} case. Colors indicate the same  cross-sections shown in the top panels. Magnitudes are in dimensionless units and logarithmic scale. Small black triangles indicate the frequency at which the maximum current is achieved. }}
    \label{fig:localHeatFluxes}
\end{figure}

In order to understand the effect of the hole on the heat transport along the plate, we first analyze the amplitude of the heat current defined in Eq.($\ref{stationaryIntegralHeatFlux}$). The position of the chosen sections are depicted in Fig. $\ref{fig:localHeatFluxes}$.

In the adiabatic limit $\omega \rightarrow 0$, the heat current amplitude converge to the same value for all positions $x$. This value corresponds to the case of a static temperature gradient, where the heat current is uniform along the plate.

For a given frequency, the value of the local heat current decreases as we move away from the border with oscillating temperature. For positions $x$ near to the constant temperature boundary, the current amplitude presents a monotonic decreasing behavior with the frequency. However for positions closer to the oscillatory temperature, the flux displays a non monotonic dependence showing a maximum amplitude for a characteristic resonant frequency $\omega^*$, indicated with black triangles in Fig $\ref{fig:localHeatFluxes}$. These resonant frequencies are sensitive to the shape and size of the hole.

In other words, geometry and frequency are parameters that can be tuned to maximize the heat current in a desired position of the structure.

\subsubsection*{Global Rectification}

Comparing the heat currents in the positions $x=0$ and $x=L$ for the the {\it left} and {\it right} cases, gives us the response of the whole system as a rectifier. We define this performance as a global rectification being computed with Eq. ($\ref{rectification}$) at $x=0$.  

\begin{figure}[ht!]
    \centering
    \includegraphics[scale=0.4]{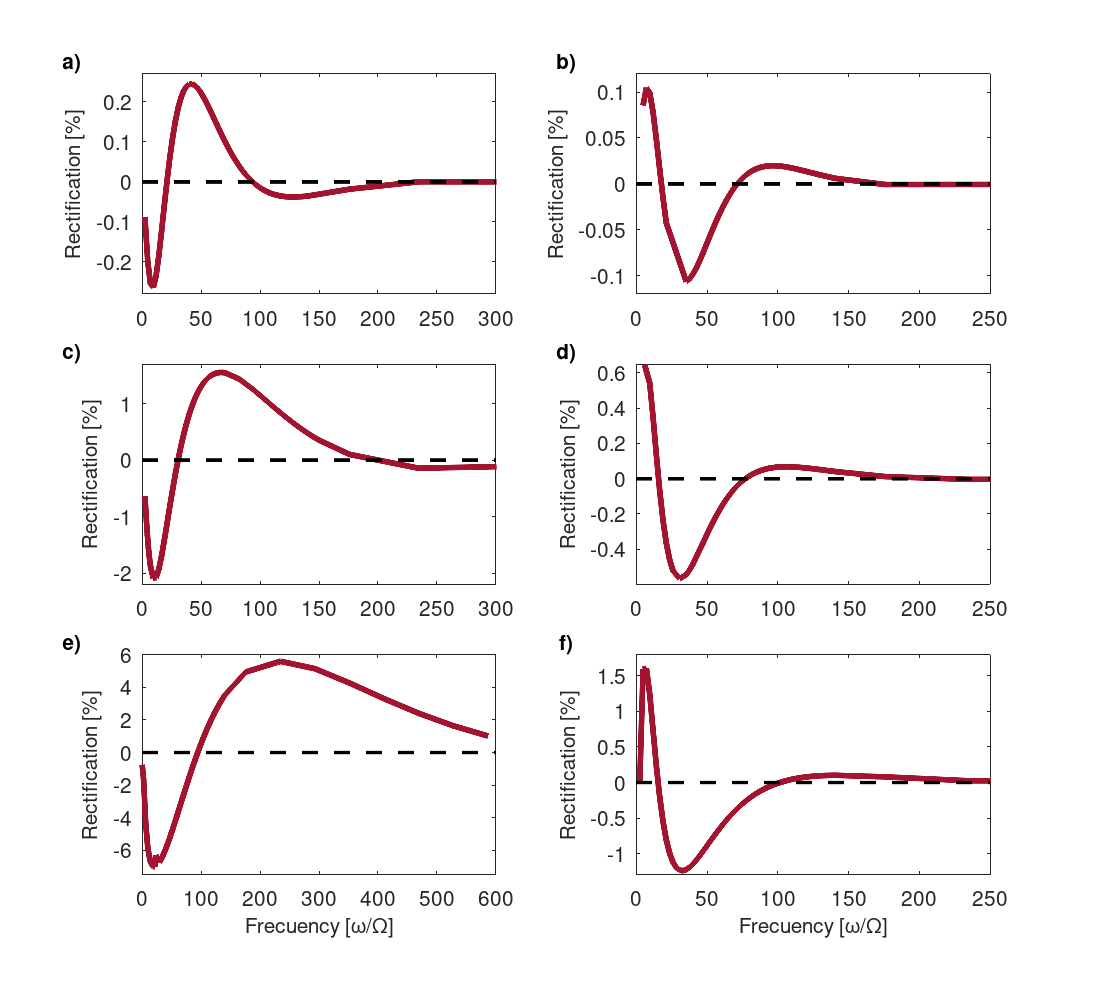}
    \caption{Global rectification as a function of $\omega$ for a triangle/crescent (left/right panels). Top panels (a) and (b) correspond to cross section $C = 0.25$, medium panels (c) and (d) $C =0.5$, low panels (e) and (f) $C = 0.75$.}
    \label{fig:globalRectification}
\end{figure}

Fig. \ref{fig:globalRectification} shows the global rectification coefficient as a function of the frequency, for different shapes and sizes of the hole. In all cases we observe a change from positive to negative values, with maxima and minima located at characteristic frequencies that depend on the geometry of the hole. This suggests a kind of resonant rectification, due to the match between temporal and spatial scales of the heat conduction from one side to the opposite one of the plate.

This inversion of the rectification was also observed in the thermal transport in quantum systems, as well as in magnetic and electronic systems \cite{zhangReversalThermalRectification2009, saRectificationIonCurrent2013, dalgleish2006a}.

For frequencies $\omega \gg \Omega$, and far away from the inversion and the maxima and minima, the rectification goes to zero for all geometries. When temperature oscillates very rapidly, the system can no longer respond accordingly to the disturbance. If the thermal penetration depth is much shorter than the distance from the thermal contact to the hole, then it becomes virtually {\it invisible}. Therefore, heat transport in both directions is identical so no rectification occurs. This is the reason why decay of the rectification takes place for larger frequencies when the hole has a larger size (see panels e) and f) in Fig. \ref{fig:globalRectification}),  compared with holes of smaller size (panels a) and b)).

For very low frequencies (adiabatic limit $\omega \ll \Omega$), the system behaves as being subjected to a static temperature gradient. The thermal current is practically the same in all sections along the plate, as it is shown in Fig. \ref{fig:localHeatFluxes}. When the oscillating bath is changed from left to right, the situation is analogous, and the heat current only depends on the temperature difference. The whole system has a constant thermal resistance given by the shape and size of the hole, and the rectification does not take place. 

\subsubsection*{Local Rectification}

\begin{figure}[ht!]
    \centering
    \includegraphics[scale=0.4]{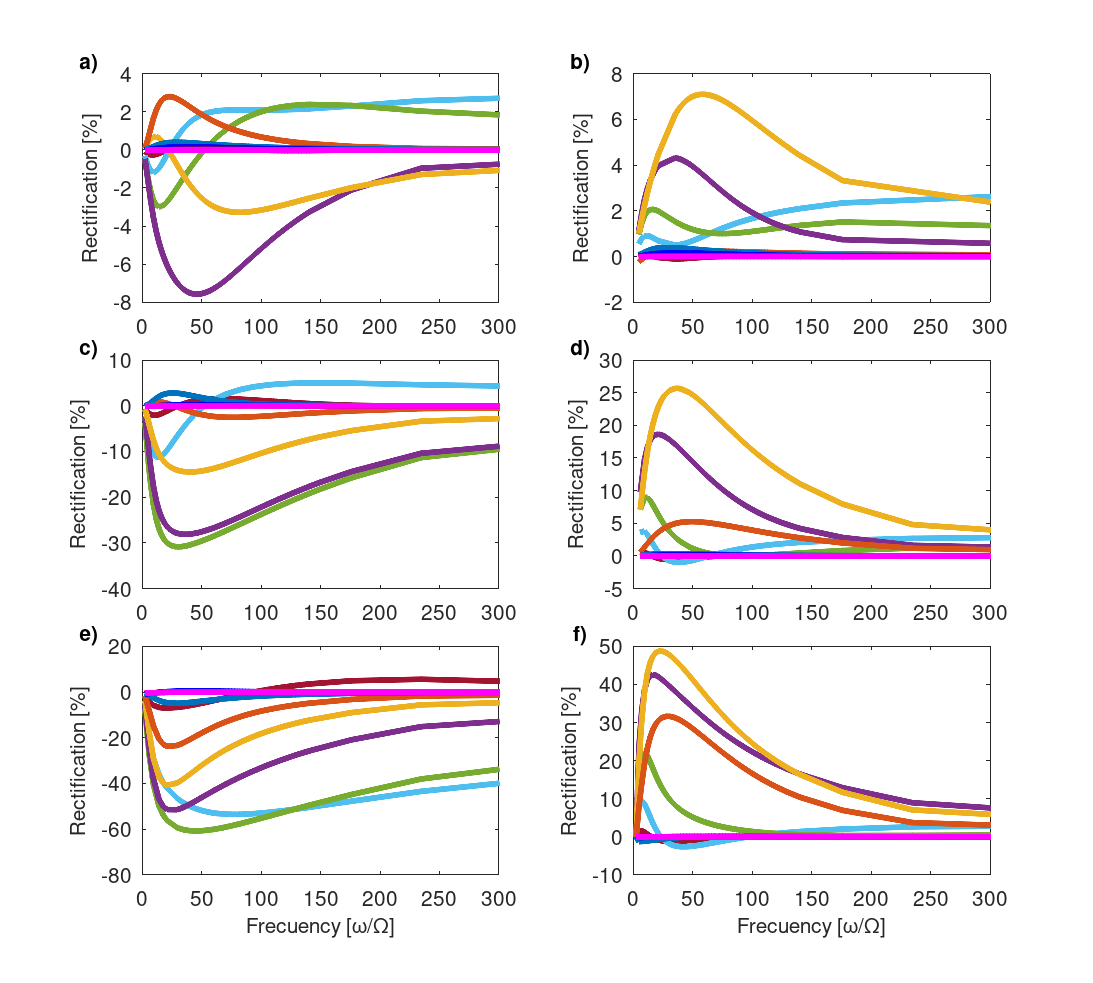}
    \caption{Local rectification as a function of $\omega$ for a triangle/crescent (left/right panels). Top panels (a) and (b) correspond to cross section $C = 0.25$, medium panels (c) and (d) $C =0.5$, low panels (e) and (f) $C = 0.75$. Color code is indicated in top panel of Fig.\ref{fig:localHeatFluxes}.}
    \label{fig:localRectification}
\end{figure}

We now analyze in Fig. \ref{fig:localRectification} the local rectification as defined in Eq. (\ref{rectification}) for different positions along the plate (same color code as in Fig. \ref{fig:localHeatFluxes}) as a function of the frequency. In order to compare with the global rectification we consider the same two hole geometries with cross sections  $C=0.25$, $C=0.50$ and $C=0.75$.
Comparing  Figs. \ref{fig:globalRectification} and \ref{fig:localRectification}, we observe that the local rectification can be substantially bigger than the global ones, with maximum values that depend sensitively on the parameters.

For a crescent geometry, the rectification values are predominantly positive for all positions and sizes. However, for the triangle, positive and negative values are observed. As the cross section of the hole increases, local negative rectification becomes dominant, as can be observed in the left column of Fig. \ref{fig:localRectification}.

It is interesting to note that for the crescent hole maximum local rectification is achieved near the middle of the plate (orange curves). However for the triangular hole this maximum is achieved at $x \approx L/3$ (violet and green curves), that is,  positions closer to the borders.

\section{Conclusions}
In this work, we introduced a 2D model of a solid plate with a hole, that can act as a thermal diode when subjected to a temperature cycling. We found that the system rectifies the heat current locally and globally, which was quantified using a rectification coefficient dependent on the position on the plate. Moreover, the rectification achieves a maximum and a minimum of different sign, being the frequency a key parameter that switches the preferred heat transport direction from left to right.

Geometry is the other relevant parameter to tune the maximum value of the rectification and the frequency at which the inversion is produced. For the analyzed shapes, the triangular hole shows a better performance as a rectifier than the crescent hole.

It is interesting to note that our device can be designed with inexpensive and easily obtained materials chosen accordingly to a desired frequency, achieving a global rectification of 10\%, or even larger if considering the local rectification. Just to give some examples we
consider typical materials used in thermal devices. A plate of 10 cm$^2$ with a triangle hole, made of Cu or Si with thermal diffusivities in the range 0.88 to 1.11 (cm$^2$/s) shows a maximum global rectification at frequencies around 50 Hz, typical of AC current. Another example is PDMS material for microfluidic devices, with a thermal diffusivity of 10$^{-3}$ (cm$^2$/s). This gives a maximum rectification at 0.06 Hz (period of 15 s), accordingly to response times of a typical microfluidic diffusor at low Reynolds number. In other words, we have proposed a very simple device that can be incorporated to dissipative thermal architectures, in order to produce a {\it dynamical} thermal diode effect assisted by geometry, in the presence of temperature cycling as found in many natural or industrial environments.

\section*{Acknowledgments}
{L.Z., L.C.,  A.M. and F.C. thank: \newline PIP-11220200101599CO and \newline FONCYT-ANPCyT-PICT-2018-02306.}

\section*{Declaration of Competing Interest}
{None}

\bibliographystyle{unsrt}  
\bibliography{main}

\end{document}